\documentclass[aps,twocolumn,showpacs,preprintnumbers,amsmath,amssymb,superscriptaddress,floatfix,nofootinbib]{revtex4}
\usepackage{lipsum}
\usepackage{graphicx}
\graphicspath{{images/}}
\usepackage{epsfig}
\usepackage{epstopdf}
\usepackage{hyperref}
\usepackage{amsmath}
\usepackage{amsfonts}
\usepackage{amssymb}
\usepackage{caption}
\usepackage{subcaption}
\DeclareMathSizes{12}{12}{12}{12}
\usepackage{wrapfig}
\usepackage[normalem]{ulem}  

\renewcommand\sout{\bgroup \color[rgb]{1,0,0} \ULdepth=-.5ex \ULset}

\numberwithin{equation}{section} 
\usepackage{tikz}
\usetikzlibrary{arrows,shapes}
\usetikzlibrary{trees}
\usetikzlibrary{matrix,arrows} 				
\usetikzlibrary{positioning}				
\usetikzlibrary{calc,through}				
\usetikzlibrary{decorations.pathreplacing}  
\usepackage{pgffor}							

\usetikzlibrary{decorations.pathmorphing}	
\usetikzlibrary{decorations.markings}
\tikzset{
vector/.style={decorate, decoration={snake}, draw},
provector/.style={decorate, decoration={snake,amplitude=2.5pt}, draw},
antivector/.style={decorate, decoration={snake,amplitude=-2.5pt}, draw},
fermion/.style={draw=black, postaction={decorate},
	decoration={markings,mark=at position .55 with {\arrow[draw=black]{>}}}},
fermionbar/.style={draw=black, postaction={decorate},
	decoration={markings,mark=at position .55 with {\arrow[draw=black]{<}}}},
fermionnoarrow/.style={draw=black},
gluon/.style={decorate, draw=black,
	decoration={coil,amplitude=4pt, segment length=5pt}},
scalar/.style={dashed,draw=black, postaction={decorate},
	decoration={markings,mark=at position .55 with {\arrow[draw=black]{>}}}},
scalarbar/.style={dashed,draw=black, postaction={decorate},
	decoration={markings,mark=at position .55 with {\arrow[draw=black]{<}}}},
scalarnoarrow/.style={dashed,draw=black},
electron/.style={draw=black, postaction={decorate},
	decoration={markings,mark=at position .55 with {\arrow[draw=black]{>}}}},
bigvector/.style={decorate, decoration={snake,amplitude=4pt}, draw},
}

\tikzstyle{block} = [draw, rectangle, 
    minimum height=3em, minimum width=6em]

\begin{document}
\title{The role of the triangle singularity in $\Lambda(1405)$
production in the $\pi^-p\rightarrow
K^0\pi\Sigma$ and $pp\rightarrow pK^+\pi\Sigma$ processes}
\date{\today}

\author{M.~Bayar}
\email{melahat.bayar@kocaeli.edu.tr}
\affiliation{Department of Physics, Kocaeli University, 41380 Izmit, Turkey}
\affiliation{Departamento de
F\'{\i}sica Te\'orica and IFIC, Centro Mixto Universidad de Valencia-CSIC Institutos de Investigaci\'on de Paterna, Aptdo.22085, 46071 Valencia, Spain}
\author{R.~Pavao}

\author{S.~Sakai}

\author{E.~Oset}
\affiliation{Departamento de
F\'{\i}sica Te\'orica and IFIC, Centro Mixto Universidad de Valencia-CSIC Institutos de Investigaci\'on de Paterna, Aptdo.22085, 46071 Valencia, Spain}

\begin{abstract} 
 We have investigated the cross section for the $\pi^-p\rightarrow
 K^0\pi\Sigma$ and $pp\rightarrow pK^+\pi\Sigma$
 reactions
 paying attention to a mechanism that develops a triangle singularity.
 The triangle diagram is realized by the decay of a $N^*$ to
 $K^*\Sigma$ and the $K^*$ decay into $\pi K$, and
 the $\pi\Sigma$ finally merges into $\Lambda(1405)$.
 The mechanism is expected to produce a
 peak around
 $2140$~MeV in the $K\Lambda(1405)$ invariant mass.
 We found that a clear peak appears around $2100$~MeV in the $K\Lambda(1405)$
 invariant mass which is about $40$~MeV
 lower
 than
 the expectation, and that is due to the resonance peak of
 a
 $N^*$ resonance which plays a crucial role in the $K^*\Sigma$
 production.
 The mechanism studied produces the peak of the $\Lambda(1405)$
 around or below 1400~MeV, as is seen in the $pp\rightarrow
 pK^+\pi\Sigma$ HADES experiment.
\end{abstract}


\maketitle

\section{Introduction}
The nature of the $\Lambda(1405)$, the lowest excitation of $\Lambda$ with
$J^P=1/2^-$, has been given much attention for a long time.
The quark model predicts the mass at higher energy than 
the observed peak \cite{Isgur:1978xj}, and
a description of the $\Lambda(1405)$ as a $\bar{K}N$ molecular state
shows a good agreement with the experimental result, as originally pointed out in
Refs.~\cite{Dalitz:1959dn,Dalitz:1960du,Dalitz:1967fp}.
The studies of the $\bar{K}N$ system based on SU(3) chiral
symmetry with the implementation of unitarity and coupled channels
suggest that the
$\Lambda(1405)$ is generated as a $\bar{K}N$ quasi-bound state
\cite{Kaiser:1995eg,Oset:1997it,Oller:2000fj,Lutz:2001yb,GarciaRecio:2002td,Jido:2003cb,Borasoy:2005ie,Ikeda:2011dx,Ikeda:2012au,Guo:2012vv,Feijoo:2015yja,Mai:2014xna}.
The recent analysis of the lattice QCD simulation
supports the molecular picture of the $\Lambda(1405)$
\cite{Hall:2014uca,Molina:2015uqp}.
Furthermore, the analysis of the compositeness
\cite{Sekihara:2014kya,Guo:2015daa,Kamiya:2015aea} which is a measure
of the hadronic molecular component,
the charge radius \cite{Sekihara:2008qk}, and the root mean square radius
\cite{Miyahara:2015bya}, also support the picture of the $\Lambda(1405)$
as a $\bar{K}N$ molecule.
Other than these works, many studies for the $\Lambda(1405)$
production from the photon-induced reaction
\cite{Nacher:1998mi,Nacher:1999ni,Borasoy:2007ku,Roca:2013av,Nakamura:2013boa,Wang:2016dtb},
the pion-induced one \cite{Magas:2005vu,Hyodo:2003jw},
the kaon-induced one
\cite{Jido:2009jf,Miyagawa:2012xz,Jido:2012cy,Ohnishi:2015iaq},
the proton-proton collision \cite{Geng:2007vm,Siebenson:2013rpa},
and the heavy meson decay \cite{Miyahara:2015cja} were
carried out to clarify the nature of the $\Lambda(1405)$ resonance.
The studies related to the $\bar{K}N$ system are summarized in
Refs.~\cite{Hyodo:2011ur,Kamiya:2016jqc} (see also note in the PDG \cite{pdg2}).

In Ref.~\cite{Wang:2016dtb}, the role of the triangle singularity (TS) on the
angle and the energy dependence of the $\Lambda(1405)$ photoproduction
was studied.
The triangle singularity was first pointed out in Ref.~\cite{Landau:1959fi}.
The corresponding Feynman diagram is formed by a sequential decay of a
hadron and a fusion of two of them, and the amplitude associated with
the diagram has a singularity if the process has a classical
counterpart, which is known as Coleman-Norton theorem
\cite{Coleman:1965xm}.
The studies of many processes including the triangle singularity elucidate
the possible effect of the triangle singularity on the hadron properties,
$e.g.$, the $\eta(1405)$ decay into $\pi^0a_0$ or $\pi^0f_0$
\cite{Wu:2011yx,Aceti:2012dj,Wu:2012pg}, the possible origin of $Z_c(3900)$
\cite{Wang:2013cya,Liu:2013vfa,Liu:2015taa}, the speculation on the pentaquark
candidate $P_c$ \cite{Guo:2015umn,Liu:2015fea,Guo:2016bkl} (see
Ref.~\cite{Bayar:2016ftu} for a critical discussion to the light of the
preferred experimental quantum numbers \cite{expe}),
the $B_s$ decay into $B\pi\pi$ \cite{Liu:2017vsf}
and $B^-$ decays \cite{Sakai:2017hpg,Pavao:2017kcr}.
Here, we note that
the strength of the triangle peak is tightly connected with the coupling
strength of
the two hadrons merging into a third one.
For example, in the study of the $B^-$ decay into
$K^-\pi^-D_{s0}^+(D_{s1}^+)$ \cite{Sakai:2017hpg}, the $DK$
($D^*K$) in the triangle loop goes into $D_{s0}$ ($D_{s1}$), which is
dynamically generated from the $DK$ ($D^*K$) and has a large coupling
to this channel
\cite{Gamermann:2006nm,Gamermann:2007fi}.
Then, the observation of the peak from the triangle mechanism would give
an additional support for the hadronic molecular picture of these
states.

\begin{figure}[t]
 \includegraphics[width=5cm]{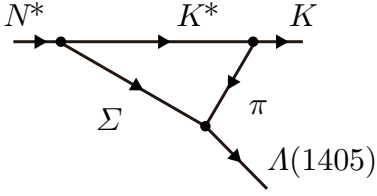}
 \caption{Triangle diagram for the $\Lambda(1405)$ production from a
 $N^*$ resonance.}
 \label{fig1}
\end{figure}

For further understanding of the nature of the $\Lambda(1405)$ and
triangle mechanisms,
in this paper we investigate the $\pi^-p\rightarrow K^0\pi\Sigma$
and $pp\rightarrow pK^+\pi\Sigma$ processes including a triangle
diagram.
In both processes, the triangle diagram is formed by a $N^*$ decay into
$K^*\Sigma$ followed by the decay of $K^*$ into $\pi K$ and the fusion of
the $\pi \Sigma$ to form the $\Lambda(1405)$, which finally decays into $\pi\Sigma$.
In this process, the $K^*\pi\Sigma$ loop generates a triangle singularity
around $2140$~MeV in the invariant mass of $K\Lambda(1405)$ from the
formula given by Eq.~(18) of Ref.~\cite{Bayar:2016ftu}.
The corresponding diagram is shown in Fig.~\ref{fig1}.
The $N^*$ resonance which strongly couples to $K^*\Sigma$ is
obtained in Ref.~\cite{Oset:2009vf} based on the hidden local
symmetry and the chiral unitary approach, and the analysis of
the $K\Sigma$ photoproduction off nucleon around the $K^*\Lambda$
threshold energy suggests that the resonance is responsible for the
observed cross section \cite{Ramos:2013wua}.

As the result of our calculation, we found a peak in the
$K\Lambda(1405)$ mass distribution around 2100~MeV in
both reactions,
which is lowered with respect to the 2140~MeV given by the TS
master formula \cite{Bayar:2016ftu} by the initial $N^*$ resonance
which peaks around 2030~MeV.
The experimental study on the $\Lambda(1405)$ production from the
$\pi^-p$ is reported in Refs.~\cite{Engler:1965zz,Thomas:1973uh},
but the energy is too small for the triangle
singularity from the $K^*\pi\Sigma$ loop to be observed.
The production of the $\Lambda(1405)$ from the proton-proton collision
is studied in Refs.~\cite{Zychor:2007gf,Agakishiev:2012xk,Adamczewski-Musch:2016vrc}.
The future observation of the inevitable peak from the triangle
mechanism induced by the $\Lambda(1405)$ would give further
support for the molecular nature of the $\Lambda(1405)$.

\section{Formalism}

\subsection{ $\mathbf{\pi^-p\rightarrow K^0 \pi \Sigma}$}

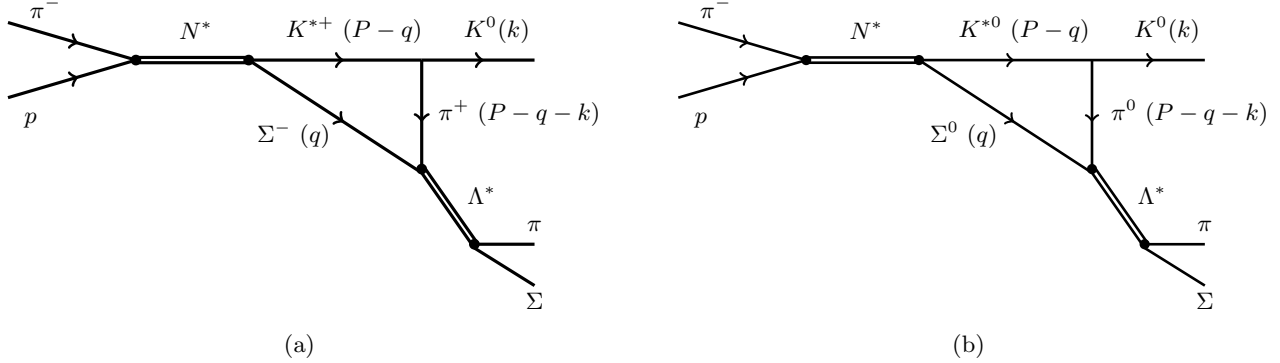
\begin{figure*}[t]
    \centering
    \begin{subfigure}[t]{0.5\textwidth}

        \centering
\begin{tikzpicture}[line width=1.2 pt]

	\draw[fermion] (-1.5,2.5) -- (0.2,2);
    \draw[fermion] (-1.5,1.5) -- (0.2,2);
    
    \draw[fermionnoarrow] (0.2,2.035) -- (1.7,2.035);
    \draw[fermionnoarrow] (0.2,1.965) -- (1.7,1.965);
    
	\draw[fermion] (1.7,2) -- (4,2);
	\draw[fermion] (4,2) -- (5.5,2);
	\draw[fermion] (4,2) -- (4,0.5);
	\draw[fermion] (1.7,2) -- (4,0.5);

	\draw[fermionnoarrow] (4,0.5) -- (4.7,-0.5);
	\draw[fermionnoarrow] (4.05,0.55) -- (4.75,-0.45);
	
	\draw[fermionnoarrow] (4.75,-0.45) -- (5.5,-0.45);
	\draw[fermionnoarrow] (4.7,-0.5) -- (5.5,-1);

    \node at (-1, 2.7) {$ \pi^-$};
    \node at (-1.2, 1.2) {$ p$};
    \node at (1, 2.4) {$ N^*$};
    \node at (3.1, 2.4) {$K^{* +}  \ (P-q)$ };
    \node at (5.3, 1.3) {$\pi^+ \ (P-q-k)$};
    \node at (2.3, 1) {$\Sigma^- \ (q)$};
    \node at (5, 2.4) {$K^0 (k)$};
    \node at (4.8, 0.2) {$\Lambda^* $};
    \node at (5.5, -0.2) {$ \pi$};
    \node at (5.5, -1.2) {$ \Sigma$};
    
    \draw[fill=black] (0.2,2) circle (.05cm);
    \draw[fill=black] (1.7,2) circle (.05cm);
    
    \draw[fill=black] (4,0.55) circle (.05cm);
    \draw[fill=black] (4.7,-0.45) circle (.05cm);

\end{tikzpicture}
        \caption{ $\ $ }
            	\label{fig::fig1a}
    \end{subfigure}%
    ~ 
    \begin{subfigure}[t]{0.5\textwidth}
        \centering

\begin{tikzpicture}[line width=1 pt]

	\draw[fermion] (-1.5,2.5) -- (0.2,2);
    \draw[fermion] (-1.5,1.5) -- (0.2,2);
    
    \draw[fermionnoarrow] (0.2,2.035) -- (1.7,2.035);
    \draw[fermionnoarrow] (0.2,1.965) -- (1.7,1.965);
    
	\draw[fermion] (1.7,2) -- (4,2);
	\draw[fermion] (4,2) -- (5.5,2);
	\draw[fermion] (4,2) -- (4,0.5);
	\draw[fermion] (1.7,2) -- (4,0.5);

	\draw[fermionnoarrow] (4,0.5) -- (4.7,-0.5);
	\draw[fermionnoarrow] (4.05,0.55) -- (4.75,-0.45);
	
	\draw[fermionnoarrow] (4.75,-0.45) -- (5.5,-0.45);
	\draw[fermionnoarrow] (4.7,-0.5) -- (5.5,-1);

    \node at (-1, 2.7) {$  \pi^-$};
    \node at (-1.2, 1.2) {$  p$};
    \node at (1, 2.4) {$  N^*$};
    \node at (3.1, 2.4) {$ K^{* 0}  \ (P-q)$ };
    \node at (5.3, 1.3) {$ \pi^0 \ (P-q-k)$};
    \node at (2.3, 1) {$ \Sigma^0 \ (q)$};
    \node at (5, 2.4) {$  K^0 (k)$};
    \node at (4.8, 0.2) {$ \Lambda^* $};
    \node at (5.5, -0.2) {$  \pi$};
    \node at (5.5, -1.2) {$   \Sigma$};
    
    \draw[fill=black] (0.2,2) circle (.05cm);
    \draw[fill=black] (1.7,2) circle (.05cm);
    
    \draw[fill=black] (4,0.55) circle (.05cm);
    \draw[fill=black] (4.7,-0.45) circle (.05cm);

\end{tikzpicture}
        \caption{ $\ $ }
               \label{fig::fig1b}
    \end{subfigure}
    \caption{Diagrams for the reaction $ \pi^- p \rightarrow K^0 \pi \Sigma$ that contain the triangle mechanism, where $\pi \Sigma$ can be $\pi^- \Sigma^+, \ \pi^0 \Sigma^0 \ \text{and} \ \pi^+ \Sigma^-$.}
    \label{fig::fig1}
\end{figure*}

In this subsection we will study the effects of the triangle loop in the
following decays: $\pi^- p \rightarrow K^0 \pi^+ \Sigma^-$, $\pi^- p
\rightarrow K^0 \pi^0 \Sigma^0$ and $\pi^- p \rightarrow K^0 \pi^-
\Sigma^+$. The diagrams where the triangle singularity can appear for
those reactions are shown in Fig.~\ref{fig::fig1}. To evaluate the
differential cross section associated with this diagram we will use
\begin{equation}
\label{eq:cross1}
\frac{d \sigma_{K^0\pi \Sigma}}{d m_{\text{inv}}} = \frac{M_p M_{\Sigma}
\ |\vec{k}| |\vec{\tilde{p}}_{\pi}|}{2 (2 \pi)^3 \ s \ |\vec{p}_{\pi}|}
\overline{\sum} \sum |t_{\pi^- p \rightarrow K^0 \pi \Sigma}|^2,
\end{equation}
with $m_{\text{inv}}$ the invariant mass of the final $\pi \Sigma$,
\begin{equation}
\label{eq:pimom}
|\vec{p}_{\pi}| = \frac{\lambda^{\frac{1}{2}}(s, m_{\pi}^2, M_N^2 )}{2 \sqrt{s}},
\end{equation}
the momentum of the initial $\pi^-$ in the $\pi^- p$ center-of-mass frame (CM),
\begin{equation}
|\vec{k}| = \frac{\lambda^{\frac{1}{2}}(s, m_{K}^2, m_{\text{inv}}^2 )}{2 \sqrt{s}},
\end{equation}
the momentum of the final $K^0$ in the $\pi^- p$ CM, and
\begin{equation}
|\vec{\tilde{p}}_{\pi}|= \frac{\lambda^{\frac{1}{2}}(m_{\text{inv}}^2,  m_{\pi}^2, M_{\Sigma}^2 )}{2 m_{\text{inv}}},\label{eqII4}
\end{equation}
the momentum of the final $\pi$ in the $\pi \Sigma$ CM.

The resonance $N^*(2030)$ could have $J^P=\frac{1}{2}^-$ or
$J^P=\frac{3}{2}^-$  as stated in Refs.~\cite{Oset:2009vf,
Ramos:2013wua} with a width $\Gamma_{N^*}\simeq 125$ MeV,
but in order to have $J=3/2$ in $\pi^-p$ we need $p$-wave and then
we have positive parity.
Hence, we take $\pi^-p$ in $L=0$ and hence $J^P_{N^*}=\frac{1}{2}^-$.
In the isospin basis the $\pi^-p\rightarrow N^\ast$ vertex has then the form
\begin{equation}
-it_{\pi N, N^* }= -i g^{I=\frac{1}{2}}_{\pi N, N^* }.
\end{equation}

To estimate the $g^{I=\frac{1}{2}}_{\pi N, N^* }$ we
assume that $\Gamma_{N^*, \pi N}$ is of the order of $70$ MeV and then
use 
{the formula,
\begin{align}
 \Gamma_{N^*, \pi N} =& \frac{1}{2 \pi} \frac{M_{\Sigma}}{M_{N^*}} (g^{I=\frac{1}{2}}_{\pi N, N^* })^2 \ |\vec{p}_{\pi}|
\end{align}
with $M_{N^*}$ the mass of $N^*(2030)$. Here, $|\vec{p}_{\pi}|$ is the
momentum of $\pi$ that results from the decay of $N^*$ and is evaluated
using Eq.~\eqref{eq:pimom}, $s=M_{N^*}^2$.
Finally, we obtain
$g^{I=\frac{1}{2}}_{\pi N, N^* } \simeq 1.1$.}

Since we will have different amplitudes if we change the charge of the
intermediate $\pi \Sigma$ particles, it is convenient to go from the
isospin basis ($\left |I, I_3 \right >$) to the charge basis. Using the
Clebsch-Gordan coefficients, we have
\begin{equation}
\left|\pi^- p \right>= \sqrt{\frac{1}{3}}\left|\frac{3}{2}, -\frac{1}{2} \right>-\sqrt{\frac{2}{3}}\left|\frac{1}{2}, -\frac{1}{2} \right>.
\end{equation}
This means that the
coupling of $\pi^- p$ to $N^*$ will be
\begin{equation}
g_{\pi^- p, N^* } =-\sqrt{\frac{2}{3}} g^{I=\frac{1}{2}}_{\pi N, N^* }
\end{equation}

The $N^*(2030) \rightarrow K^* \Sigma$ process occurs in
$s$-wave, then the amplitude is written as
\begin{equation}
- i t_{N^*, K^* \Sigma} = -i g_{N^*, \Sigma K^*} \ \vec{\sigma} \cdot \vec{\epsilon}_{K^*}.
\end{equation}
From Ref.~\cite{Oset:2009vf}, we get $g^{I=\frac{1}{2}}_{N^*, K^* \Sigma}
= 3.9 + i 0.2$, and since we
have both $\Sigma^- K^{* +} $ and $\Sigma^0 K^{* 0} $ (Figs.~\ref{fig::fig1a} and~\ref{fig::fig1b} respectively), then, using
\begin{subequations}
\begin{align}
\label{eq:CGCa}
& \left|\Sigma^0K^{* 0} \right>= \sqrt{\frac{2}{3}}\left|\frac{3}{2}, -\frac{1}{2} \right>+\sqrt{\frac{1}{3}}\left|\frac{1}{2}, -\frac{1}{2} \right>, \\
\label{eq:CGCb}
& \left|\Sigma^-K^{* +} \right>= \sqrt{\frac{1}{3}}\left|\frac{3}{2}, -\frac{1}{2} \right>-\sqrt{\frac{2}{3}}\left|\frac{1}{2}, -\frac{1}{2} \right>,
\end{align}
\end{subequations}
we get
\begin{subequations}
\begin{align}
& g_{N^*,\Sigma^0 K^{* 0}}= \sqrt{\frac{1}{3}} g^{I=\frac{1}{2}}_{N^*,
 \Sigma K^*},\\
& g_{N^*,\Sigma^- K^{* +}}= -\sqrt{\frac{2}{3}}
 g^{I=\frac{1}{2}}_{N^*,\Sigma K^*}.
\end{align}
\end{subequations}
Then, for the amplitude of the $\pi^- p \rightarrow  \Sigma K^*$
reaction
through $N^*(2030)$, we have
\begin{subequations}
\begin{align}
& t_{\pi^- p, \Sigma^0K^{* 0}} = \frac{g_{\pi^- p, N^* } \
 g_{N^*,\Sigma^0 K^{* 0}}}{\sqrt{s}-M_{N^*}+i \frac{\Gamma_{N^*}}{2}}\vec{\sigma}\cdot\vec{\epsilon}_{K^*}, \\
& t_{\pi^- p, \Sigma^-K^{* +}} =\frac{g_{\pi^- p, N^* } \
 g_{N^*,\Sigma^- K^{* +}}}{\sqrt{s}-M_{N^*}+i \frac{\Gamma_{N^*}}{2}}\vec{\sigma}\cdot\vec{\epsilon}_{K^*}.
\end{align}
\end{subequations}

Now, the $K^{* +} \rightarrow K^0 \pi^+$ vertex can be calculated using the chiral invariant Lagrangian with local hidden symmetry given in Refs.~\cite{LHS1,LHS2,LHS3,LHS4},
\begin{equation}
\label{eq:vpp}
\mathcal{L}_{VPP} = -i g \left <  \left[\Phi, \partial_{\mu} \Phi\right] V^{\mu} \right >.
\end{equation}
 The symbol $\left<...\right>$ here represents the trace over the SU(3) flavor matrices, and the coupling is $g=m_V/2 f_{\pi}$, with $m_V=800 \ \text{MeV}$ and $f_{\pi}=93 \ \text{MeV}$.  The SU(3) matrices for the pseudoscalar and vector octet mesons $\Phi$ and $V^{\mu}$ are given by
\begin{align}
\Phi=\begin{pmatrix}
\frac{1}{\sqrt{2}} \pi^0 + \frac{1}{\sqrt{6}} \eta  & \pi^+ &  K^+ \\
\pi^- & -\frac{1}{\sqrt{2}} \pi^0 + \frac{1}{\sqrt{6}} \eta & K^0\\
K^- & \bar{K}^0 & -\sqrt{\frac{2}{3}} \eta
     \end{pmatrix},
\end{align}
\begin{align}
V_{\mu}=\begin{pmatrix}
\frac{1}{\sqrt{2}} \rho^0_{\mu} + \frac{1}{\sqrt{2}} \omega_{\mu} & \rho^+_{\mu} & K^{* +}_{\mu} \\ 
\rho^-_{\mu} & -\frac{1}{\sqrt{2}} \rho^0_{\mu} + \frac{1}{\sqrt{2}} \omega_{\mu} & K^{* 0}_{\mu} \\ 
K^{* -}_{\mu} & \bar{K}^{* 0}_{\mu} & \phi_{\mu}
\end{pmatrix}.
\end{align}
From Eq.~\eqref{eq:vpp} we get
\begin{align}
\label{eq:thisone}
-i t_{K^{* +}, K^0 \pi^+} =& -i g \ \epsilon^{\mu}_{K^*}
 (p_{K^0}-p_{\pi^+})_\mu \\
 & = i g \ \epsilon^{\mu}_{K^*} (P-q-2k)_\mu \\
 & \simeq  i g \ \vec{\epsilon}_{K^*} \cdot (\vec{q}+2\vec{k}),\label{eq_kskpi}
\end{align}
where in the last step we made a nonrelativistic approximation,
neglecting the $\epsilon^0_{K^*}$ component.
This is very accurate when the momentum of the $K^*$ is small compared
to its mass. We shall evaluate the triangle diagram in the $\Sigma K^*$ CM,
where the on-shell momentum of the $K^*$ is about $250$~MeV$/c$ at
$M_{\rm inv}(\Sigma K^*)\simeq 2140$~MeV where the triangle singularity appears.
In Ref.~\cite{Sakai:2017hpg} it is shown that the effect of neglecting
the $\epsilon^0$ component goes as $(p_{K^*}/m_{K^*})^2$, with a
coefficient in front that renders this correction negligible.

Similarly, for $K^{* 0} \rightarrow K^0 \pi^0$ we get
\begin{equation}
\label{eq:thistwo}
-i t_{K^{* 0}, K^0 \pi^0} =  -i \frac{1}{\sqrt{2}} g \ \vec{\epsilon}_{K^*} \cdot (\vec{q}+2\vec{k}).
\end{equation}

The final vertex that we need to calculate in the diagrams of Fig.1 is
$t_{\Sigma \pi, \Sigma \pi}$ which is given by the $\Sigma \pi
\rightarrow \Lambda (1405) \rightarrow \Sigma \pi$
amplitude studied in Ref.~\cite{Oset:1997it} based on chiral unitary
approach.
There, the authors use the lowest order meson-baryon chiral lagrangian
\begin{equation}
\mathcal{L}_1^{(B)} = \left< \bar{B} i \gamma^{\mu} \frac{1}{4 f^2} [(\Phi \partial_{\mu} \Phi - \partial_{\mu} \Phi \Phi)B-B(\Phi \partial_{\mu} \Phi - \partial_{\mu} \Phi \Phi)] \right>,
\end{equation}
with,
\begin{equation}
B=\begin{pmatrix}
\frac{1}{\sqrt{2}} \Sigma^0 + \frac{1}{\sqrt{6}} \Lambda  & \Sigma^+ & p\\
\Sigma^- & -\frac{1}{\sqrt{2}} \Sigma^0 + \frac{1}{\sqrt{6}} \Lambda  & n\\
\Xi^- & \Xi^0  & -\sqrt{\frac{2}{3}} \Lambda
\end{pmatrix}.
\end{equation}
The
Bethe-Salpeter
equation is then used to calculate the meson-baryon amplitude,
\begin{align}
t=[1-V G]^{-1} V, \label{eq:LSeq}
\end{align}
where $t$, $V$, and $G$ are the meson-baryon amplitude, interaction kernel, and meson-baryon loop function, respectively.
{For the evaluation of $t$, we use the momentum cutoff $q_{\rm
max}=630$~MeV for the loop function $G$, and $f=1.15f_\pi$ with the pion
decay constant $f_\pi=93$~MeV as done in Ref.~\cite{Oset:1997it}.}

Thus, the amplitude associated with the diagram in
Fig.~\ref{fig::fig1a}, that we call $t_1$,
is given by
\begin{widetext}
\begin{equation}
\label{eq:t_total0}
t_1 = - i \frac{2}{3} \ \frac{g^{I=\frac{1}{2}}_{\pi N, N^* } \
g^{I=\frac{1}{2}}_{N^*, \Sigma K^*}}{\sqrt{s}-M_{N^*}+i \frac{\Gamma_{N^*}}{2}} \ g \sum_{\text{pol. of } K^* } \int \frac{d^4 q}{(2 \pi)^4}  \frac{2 M_{\Sigma} \ \vec{\sigma} \cdot \vec{\epsilon}_{K^*}}{q^2-M_{\Sigma}^2+i \epsilon} \frac{(2\vec{k}+\vec{q}) \cdot \vec{\epsilon}_{K^*}}{(P-q)^2-m_{K^*}^2+i \epsilon} \frac{t_{\Sigma^- \pi^+, \Sigma \pi}}{(P-q-k)^2 - m_{\pi}^2+i\epsilon}.
\end{equation}
\end{widetext}
Using the following property,
\begin{equation*}
\int d^3 q q_i f(\vec{q}, \vec{k}) = k_i \int d^3 q \frac{\vec{q} \cdot \vec{k}}{|\vec{k}|^2} f( \vec{q}, \vec{k})
\end{equation*}
with $f(\vec{q},\vec{k})$ the three propagators in the integrand of Eq.~\eqref{eq:t_total0},
and
{the formula in the nonrelativistic approximation,}
\begin{equation*}
 \sum_{\text{pol.}} {\epsilon_{K^*}}_i {\epsilon_{K^*}}_j = \delta_{ij},
\end{equation*}
Eq.~\eqref{eq:t_total0} becomes
\begin{equation}
t_1=- \frac{4 M_{\Sigma}}{3} \ \frac{g^{I=\frac{1}{2}}_{\pi N, N^* } \
 g^{I=\frac{1}{2}}_{N^*, \Sigma K^*}}{\sqrt{s}-M_{N^*}+i
 \frac{\Gamma_{N^*}}{2}} \ g \ \vec{\sigma} \cdot \vec{k} \ t_T	 \
 t_{\Sigma^- \pi^+, \Sigma \pi},
\end{equation}
with
\begin{widetext}
\begin{equation}
\label{eq:tT}
t_T=i \int \frac{d^4 q}{(2 \pi)^4}  \ \left(2+ \frac{\vec{q} \cdot \vec{k}}{|\vec{k}|^2}\right) \frac{1}{q^2-M_{\Sigma}^2+i \epsilon} \frac{1}{(P-q)^2-m^2_{K^*}+i \epsilon} \frac{1}{(P-q-k)^2 - m_{\pi}^2+i \epsilon}.
\end{equation}
\end{widetext}
Integrating $t_T$ over $q^0$, we get~\cite{Bayar:2016ftu,aceti_dias},
\begin{widetext}
\begin{align}
 t_T =& \int \frac{d^3 q}{(2 \pi)^3}   \left(2+ \frac{\vec{q} \cdot \vec{k}}{|\vec{k}|^2}\right) \frac{1}{8 \omega^* \omega \omega'} \frac{1}{k^0-\omega'-\omega^*} \frac{1}{P^0+\omega+\omega'-k^0} \frac{1}{P^0-\omega-\omega'-k^0 + i \epsilon} \times\notag\\
\label{eq:tsing_int}
&\times \frac{\{2P^0 \omega + 2 k^0 \omega' -2[\omega+\omega'][\omega+\omega'+\omega^*] \}}{P^0-\omega^*-\omega+i\epsilon},
\end{align}
\end{widetext}
where $P^0 = \sqrt{s}$, $\omega^*(\vec{q} \, )=\sqrt{m^2_{K^{* 0}}
+|\vec{q} \, |^2}$, $\omega'(\vec{q} \, )=\sqrt{m^2_{\pi}
+|\vec{q}+\vec{k}|^2}$ and  $\omega(\vec{q} \, )=\sqrt{M_{\Sigma}^2 +|\vec{q}
\, |^2}$. We regularize the integral in Eq.~\eqref{eq:tsing_int} by
using the same cutoff of the meson loop in Eq.~\eqref{eq:LSeq},
$\theta(q_{\text{max}}-|\vec{q}^{\ *}|)$, where $\vec{q}^{\ *}$ is the
$\Sigma$ momentum in the final $\pi \Sigma$ CM \cite{Bayar:2016ftu} and $q_{\text{max}}=630$ MeV.
The width of $K^*$ is taken into account by replacing $\omega^*$ with
$\omega^* - i \frac{\Gamma_{K^*}}{2}$.

For the case when $N^* (2030) \rightarrow K^{*+} \Sigma^-$, $t_2$, we have
\begin{equation}
t_2=- \frac{2 M_{\Sigma}}{3} \ \frac{g^{I=\frac{1}{2}}_{\pi N, N^* } \
 g^{I=\frac{1}{2}}_{N^*, \Sigma K^*}}{\sqrt{s}-M_{N^*}+i \frac{\Gamma_{N^*}}{2}} \ g \ \vec{\sigma} \cdot \vec{k} \ t_T	 \ t_{\Sigma^0 \pi^0, \Sigma \pi}.
\end{equation}
Thus, the total amplitude in Eq.~\eqref{eq:cross1} associated with $\pi^- p \rightarrow K^0 \pi \Sigma$ becomes
\begin{equation}
\label{eq:t1plust2}
t_{\pi^- p \rightarrow K^0 \pi \Sigma} = t_1+t_2=C \ \vec{\sigma} \cdot \vec{k} \ t_T	 ( \ t_{\Sigma^- \pi^+, \Sigma \pi}+ \frac{1}{2}  t_{\Sigma^0 \pi^0, \Sigma \pi}),
\end{equation}
with
\begin{equation}
C=-\frac{2}{3} \ g^{I=\frac{1}{2}}_{\pi N, N^* } \
 g^{I=\frac{1}{2}}_{N^*, \Sigma K^*} \ g \ \frac{2 M_{\Sigma}}{\sqrt{s}-M_{N^*}+i \frac{\Gamma_{N^*}}{2}}.
\end{equation}
Here, the $t_T$ associated with the diagrams in Figs.~\ref{fig::fig1a}
and~\ref{fig::fig1b} are the same because we use the isospin averaged
mass and width of the hadrons in $t_T$.

Calculating the square of the amplitude and summing and averaging over the spins we get
\begin{widetext}
\begin{equation}
\label{eq:t_finals1}
\overline{\sum} \sum |t_{\pi^- p \rightarrow K^0 \pi \Sigma}|^2 = |C|^2 |\vec{k}|^2 |t_T|^2 |\ t_{\Sigma^- \pi^+, \Sigma \pi}+ \frac{1}{2}  t_{\Sigma^0 \pi^0, \Sigma \pi}|^2
\end{equation}
\end{widetext}
Finally, using Eq.~\eqref{eq:t_finals1} in Eq.~\eqref{eq:cross1} we can
calculate the $\frac{d \sigma_{K^0\pi \Sigma}}{d m_{\text{inv}}}$
associated with the diagrams in Fig.~\ref{fig::fig1}.

\subsection{pp $\mathbf{\rightarrow}$ p$\mathbf{K^+ \pi \Sigma}$}

\begin{figure}[h]
\begin{tikzpicture}[line width=1.2 pt]
	\draw[fermionnoarrow] (-1.5,3) -- (4,3);
    \draw[fermionnoarrow] (-1.5,1.5) -- (0.8,1.5);
	\draw[fermionnoarrow] (0.8,3) -- (0.8,1.5);
	\draw[fermionnoarrow] (0.8,1.54) -- (2.3,1.54);
    \draw[fermionnoarrow] (0.8,1.46) -- (2.3,1.46);
	
	\draw[fermionnoarrow] (2.3,1.5) -- (4,2);
	\draw[fermionnoarrow] (2.3,1.5) -- (4,1);
	
    \draw[fill=black] (2.3,1.5) circle (.05cm);
    \draw[fill=black] (0.8,1.5) circle (.05cm);
    
    \node at (-1.2, 3.3) {$a$};   
    \node at (-1.2, 1.8) {$b$};

    \node at (1.5, 1.8) {$R$};   

    \node at (4, 3.3) {$1$};   
    \node at (4, 2.25) {$2$};
    \node at (4, 1.25) {$3$};

\end{tikzpicture}
\caption{Diagram of $a \ b \rightarrow 1 \ R \rightarrow 1 \ 2 \ 3$.}
\label{fig::simp}
\end{figure}
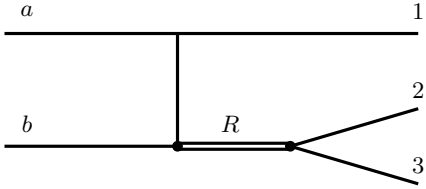

Now we will study the effects of the triangle loop in the following
decays: $p p \rightarrow pK^+ \pi^+ \Sigma^-$, $p p \rightarrow pK^+
\pi^0 \Sigma^0$ and $p p \rightarrow pK^+ \pi^- \Sigma^+$. For this, we
will first start analysing the diagram in Fig.~\ref{fig::simp}. For this
diagram, the differential cross section is calculated using
{the formula
in Ref.~\cite{Debastiani:2016xgg}},
\begin{equation}
\label{eq:dcross2}
\frac{d^2 \sigma}{dt dM_{\text{inv}}}=\frac{\prod_F \ (2 M_F)}{32 p_a^2 s (2 \pi)^3} |\vec{\tilde{p}}_2|  \overline{\sum} \sum |t_{a b \rightarrow 1 2 3}|^2,
\end{equation}
with $t=(p_a-p_1)^2$, $M_{\text{inv}}$ the invariant mass of
particles $2$ and $3$, $\vec{\tilde{p}}_2$ the momentum of particle $2$
in the $23$ CM, such that
\begin{equation}
|\vec{\tilde{p}}_2| = \frac{\lambda^{\frac{1}{2}} (M_{\text{inv}}^2, m_2^2, m_3^2)}{2 M_{\text{inv}}},
\end{equation} 
$p_a$ the momentum of the particle $a$ in the initial state,
\begin{align}
 p_a=\frac{\lambda^{\frac{1}{2}}(s,M_a^2,M_b^2)}{2\sqrt{s}},
\end{align}
and $\Pi_F \ (2 M_F)$ means that we multiply $2 M_F$ for each fermion in
Fig.~\ref{fig::simp}, where $M_F$ is the mass of the respective
fermion. This
factor
appears because we use the normalization of Ref.~\cite{mandl}.

\begin{figure*}

    \centering
    \begin{subfigure}{0.5\textwidth}
        \centering
\begin{tikzpicture}[line width=1.2 pt]

	\draw[fermion] (-1.5,3.2) -- (0.2,3.2);
    \draw[fermion] (0.2,3.2) -- (5.5,3.2);
    \draw[fermion] (-1.5,2) -- (0.2,2);
	\draw[fermionnoarrow] (0.2,3.2) -- (0.2,2);
    
    \draw[fermionnoarrow] (0.2,2.035) -- (1.7,2.035);
    \draw[fermionnoarrow] (0.2,1.965) -- (1.7,1.965);
    
	\draw[fermion] (1.7,2) -- (4,2);
	\draw[fermion] (4,2) -- (5.5,2);
	\draw[fermion] (4,2) -- (4,0.5);
	\draw[fermion] (1.7,2) -- (4,0.5);

	\draw[fermionnoarrow] (4,0.5) -- (4.7,-0.5);
	\draw[fermionnoarrow] (4.05,0.55) -- (4.75,-0.45);
	
	\draw[fermionnoarrow] (4.75,-0.45) -- (5.5,-0.45);
	\draw[fermionnoarrow] (4.7,-0.5) -- (5.5,-1);

    \node at (-1.2, 3.4) {$p$};
    \node at (-1.2, 2.2) {$p$};
    \node at (1, 2.3) {$  N^*$};
    \node at (3.1, 2.3) {$  K^{* 0}$ };
    \node at (4.5, 1.3) {$  \pi^-$};
    \node at (2.3, 1) {$  \Sigma^+$};
    \node at (5, 2.3) {$  K^+$};
    \node at (4.8, 0.2) {$  \Lambda^* $};
    \node at (5.5, -0.2) {$  \pi$};
    \node at (5.5, -1.2) {$   \Sigma$};
    \node at (5.5, 3.4) {$   p$};
    
    \draw[fill=black] (0.2,2) circle (.05cm);
    \draw[fill=black] (1.7,2) circle (.05cm);
    
    \draw[fill=black] (4,0.55) circle (.05cm);
    \draw[fill=black] (4.7,-0.45) circle (.05cm);
\end{tikzpicture}
        \caption{ $\ $ }
    \end{subfigure}%
    ~ 
    \begin{subfigure}{0.5\textwidth}
        \centering
\begin{tikzpicture}[line width=1 pt]
	\draw[fermion] (-1.5,3.2) -- (0.2,3.2);
    \draw[fermion] (0.2,3.2) -- (5.5,3.2);
    \draw[fermion] (-1.5,2) -- (0.2,2);
	\draw[fermionnoarrow] (0.2,3.2) -- (0.2,2);
    
    \draw[fermionnoarrow] (0.2,2.035) -- (1.7,2.035);
    \draw[fermionnoarrow] (0.2,1.965) -- (1.7,1.965);
    
	\draw[fermion] (1.7,2) -- (4,2);
	\draw[fermion] (4,2) -- (5.5,2);
	\draw[fermion] (4,2) -- (4,0.5);
	\draw[fermion] (1.7,2) -- (4,0.5);

	\draw[fermionnoarrow] (4,0.5) -- (4.7,-0.5);
	\draw[fermionnoarrow] (4.05,0.55) -- (4.75,-0.45);
	
	\draw[fermionnoarrow] (4.75,-0.45) -- (5.5,-0.45);
	\draw[fermionnoarrow] (4.7,-0.5) -- (5.5,-1);

    \node at (-1.2, 3.4) {$  p$};
    \node at (-1.2, 2.2) {$  p$};
    \node at (1, 2.3) {$   N^*$};
    \node at (3.1, 2.3) {$  K^{* +}$ };
    \node at (4.5, 1.3) {$  \pi^0$};
    \node at (2.3, 1) {$  \Sigma^0$};
    \node at (5, 2.3) {$  K^+$};
    \node at (4.8, 0.2) {$  \Lambda^* $};
    \node at (5.5, -0.2) {$  \pi$};
    \node at (5.5, -1.2) {$   \Sigma$};
    \node at (5.5, 3.4) {$   p$};
    
    \draw[fill=black] (0.2,2) circle (.05cm);
    \draw[fill=black] (1.7,2) circle (.05cm);
    
    \draw[fill=black] (4,0.55) circle (.05cm);
    \draw[fill=black] (4.7,-0.45) circle (.05cm);

\end{tikzpicture}
        \caption{ $\ $ }
    \end{subfigure}
    \caption{Diagrams for the reaction $ p p \rightarrow pK^+ \pi \Sigma$ that contain the triangle mechanism, where $\pi \Sigma$ can be $\pi^- \Sigma^+, \ \pi^0 \Sigma^0 \ \text{and} \ \pi^+ \Sigma^-$.}
            \label{fig::fig3}
\end{figure*}
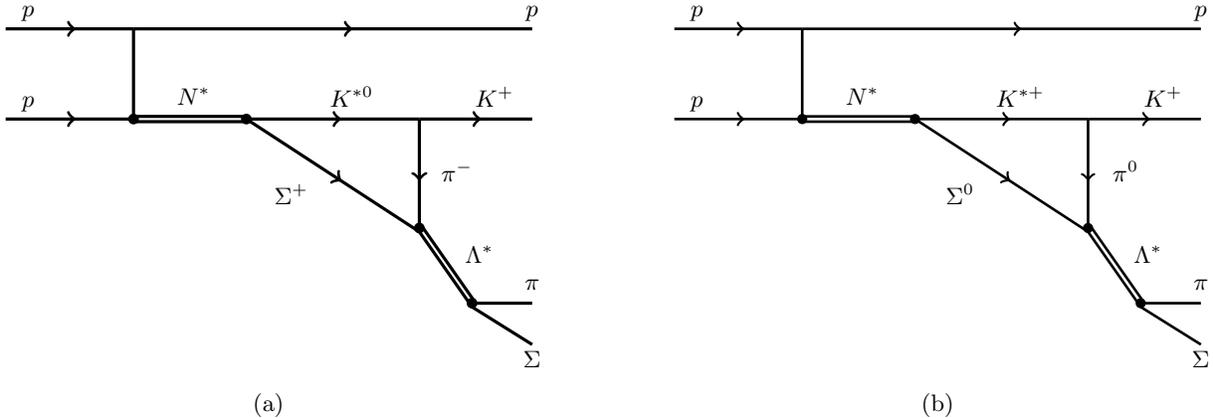

The complete diagrams for our reaction are shown in
Fig.~\ref{fig::fig3}. The triangle part of the diagrams is very similar
to the last case, except that because of charge conservation the
particles in the loop will be different.
Thus, instead of Eqs.~\eqref{eq:CGCa} and~\eqref{eq:CGCb}, we will have
\begin{subequations}
\begin{align}
& \left|\Sigma^+K^{* 0} \right>= -\sqrt{\frac{1}{3}}\left|\frac{3}{2}, \frac{1}{2} \right>-\sqrt{\frac{2}{3}}\left|\frac{1}{2}, \frac{1}{2} \right>, \\
& \left|\Sigma^0K^{* +}\right>= \sqrt{\frac{2}{3}}\left|\frac{3}{2}, \frac{1}{2} \right>-\sqrt{\frac{1}{3}}\left|\frac{1}{2}, \frac{1}{2} \right>,
\end{align}
\end{subequations}
where, to match the sign convention of the $\Phi$ and $B$ matrices, we
used $\left|\Sigma^+ \right>=-\left|1 \ 1 \right>$ (see
Ref.~\cite{Pavao:2017cpt} for further discussion).
Then, we get
$g_{N^*,\Sigma^+K^{*0}}=-\sqrt{2/3}g_{N^*,\Sigma K^*}^{I=\frac{1}{2}}$
and $g_{N^*,\Sigma^0K^{*+}}=-\sqrt{1/3}g_{N^*,\Sigma K^*}^{I=\frac{1}{2}}$.

The vertices $K^{* 0} \rightarrow K^+ \pi^-$ and $K^{* +} \rightarrow K^+ \pi^0$ are calculated using Eq.~\eqref{eq:vpp}, which gives
\begin{subequations}
\begin{align}
& -i t_{K^{* 0}, K^+ \pi^-} = i g (\vec{q}+2 \vec{k}) \cdot \vec{\epsilon}_{K^*},\\
& -i t_{K^{* +}, K^+ \pi^0} = i \frac{g}{\sqrt{2}} (\vec{q}+2 \vec{k}) \cdot \vec{\epsilon}_{K^*}
\end{align}
\end{subequations}

To calculate the cross section for the diagrams in Fig.~\ref{fig::fig3},
we proceed as done in Ref.~\cite{Debastiani:2016xgg}.
In Fig.~\ref{fig::simp}, the $t$ matrix
found in Eq.~\eqref{eq:dcross2} is given by
\begin{equation}
t_{a b \rightarrow 1 2 3} = C' \frac{1}{M_{\text{inv}}-M_R + i \frac{\Gamma_R}{2}} g_{R,23},
\end{equation}
with
$C'$ a parameter that carries the dependence of the amplitude on the
variable $t$ as well as information about the $p p \rightarrow p R$
transition.
Substituting
\begin{equation}
\Gamma_{R,23}=\frac{1}{2 \pi}  \frac{M_{3}}{M_{\text{inv}}}
 g_{R,23}^2|\vec{\tilde{p}}_2|,
\end{equation}
where particle 3 is assumed to be a baryon,
into Eq.~\eqref{eq:dcross2}, we get
\begin{widetext}
\begin{equation}
\label{eq:dcross3}
\frac{d^2 \sigma}{dt dM_{\text{inv}}}=\frac{\prod_F(2M_F)}{32 p_a^2 s (2 \pi)^3} |C'|^2 \left|\frac{1}{M_{\text{inv}}-M_R + i \frac{\Gamma_R}{2}}\right|^2 2 \pi \frac{M_{\text{inv}}}{M_{3}} \Gamma_{R,23}.
\end{equation}
\end{widetext}
Now we can take into account the complete reaction by substituting $\Gamma_{R,23}$ for $\Gamma_{N^* \rightarrow K^+ \pi \Sigma}$, where
\begin{equation}
\frac{d \Gamma_{N^* \rightarrow K^+ \pi \Sigma}}{d m_{\text{inv}}}= \frac{2 M_{N^*} 2 M_{\Sigma}}{(2 \pi)^3 4 M_{\text{inv}}^2}  |\vec{p}_K| |\vec{\tilde{p}}_{\pi}| \overline{\sum} \sum |t'|^2,
\end{equation}
with $|\vec{p}_K|$ the momentum of $K$ in the rest frame of $N^*$,
\begin{equation}
|\vec{p}_K| = \frac{\lambda^{\frac{1}{2}}(M_{\text{inv}}^2, m_K^2, m_{\text{inv}}^2)}{2 M_{\text{inv}}},
\end{equation}
and $|\vec{\tilde{p}}_{\pi}|$ the $\pi$ momentum in the $\pi\Sigma$ CM
given by Eq.~\eqref{eqII4}.

Then, from Eq.~\eqref{eq:dcross3}
we obtain
\begin{widetext}
\begin{equation}
\label{eq:dcross4}
\frac{d^3 \sigma_{pK^+\pi\Sigma}}{dt \ dM_{\text{inv}} d m_{\text{inv}}}=\frac{(2 M_p)^3 \ 2 M_{\Sigma}}{32 p_a^2 s (2 \pi)^5} |\vec{p}_K| |\vec{\tilde{p}}_{\pi}| |C'|^2 \left|\frac{1}{M_{\text{inv}}-M_{N^*} + i \frac{\Gamma_{N^*}}{2}}\right|^2  \overline{\sum} \sum |t'|^2.
\end{equation}
\end{widetext}
The transition amplitude $t'$ in Eq.~\eqref{eq:dcross4} is
\begin{equation}
\label{eq:finalt}
t'= \sqrt{\frac{2}{3}} 2M_\Sigma g^{I=\frac{1}{2}}_{N^*, \Sigma K^* } \ g \ ( t_{\Sigma^+ \pi^-, \Sigma \pi}+ \frac{1}{2}  t_{\Sigma^0 \pi^0, \Sigma \pi}) \ \vec{\sigma} \cdot \vec{k} \ t_T,
\end{equation}
which is constructed in a similar way to what was done in the previous
subsection to obtain Eq.~\eqref{eq:t1plust2} but now changing the
following variables in Eq.~\eqref{eq:tsing_int},
\begin{subequations}
\begin{align}
& P^0 = M_{\text{inv}}\\
& |\vec{k}|=|\vec{p}_K| = \frac{\lambda^{\frac{1}{2}}(M_{\text{inv}}^2,m_K^2, m_{\text{inv}}^2)}{2 M_{\text{inv}}}\\
& k^0 = \frac{M_{\text{inv}}^2 + m_K^2-m_{\text{inv}}^2}{2 M_{\text{inv}}}
\end{align}
\end{subequations}
Putting Eq.~\eqref{eq:finalt} into Eq.~\eqref{eq:dcross4} we get
\begin{widetext}
\begin{equation}
\frac{d^3 \sigma_{pK^+\pi\Sigma}}{dt \ dM_{\text{inv}} d m_{\text{inv}}}= C'' \frac{1}{|M_{\text{inv}}-M_{N^*} + i \frac{\Gamma_{N^*}}{2}|^2} |\vec{\tilde{p}}_{\pi}| |t_{\Sigma^+ \pi^-, \Sigma \pi}+ \frac{1}{2}  t_{\Sigma^0 \pi^0, \Sigma \pi}|^2 |\vec{k}|^3 |t_T|^2,
\end{equation}
\end{widetext}
where $|\vec{\tilde{p}}_{\pi}|$ is the $\pi$ momentum in the $\pi \Sigma$ CM,
\begin{equation}
|\vec{\tilde{p}}_{\pi}| = \frac{\lambda^{\frac{1}{2}}( m_{\text{inv}}^2, m_{\pi}^2, M_{\Sigma}^2)}{2  m_{\text{inv}}},
\end{equation}
and
\begin{equation}
C''= \frac{2}{3}  \frac{(2 M_p)^3 \ 2 M_{\Sigma}}{32 p_a^2 s (2
 \pi)^5}
 |g^{I=\frac{1}{2}}_{N^*,\Sigma  K^*} |^2 \ g^2 (2M_\Sigma)^2 |C'|^2,
\end{equation}
which is a function of $s=(p_a+p_b)^2$ and $t=(p_a-p_1)^2$.

Using now the relation
\begin{equation}
dt = 2 |\vec{p}_a| |\vec{p}_1| \ d \cos \theta,
\end{equation}
which follows from $t = (p_a-p_1)^2$, then we obtain
\begin{widetext}
\begin{equation}
\label{eq:finalmassint}
\frac{d^3 \sigma_{pK^+\pi\Sigma}}{d \cos \theta \ dM_{\text{inv}} d m_{\text{inv}}}=C'' \frac{2 |\vec{p}_a| |\vec{p}_1|}{|M_{\text{inv}}-M_{N^*} + i \frac{\Gamma_{N^*}}{2}|^2} |\vec{\tilde{p}}_{\pi}| |t_{\Sigma^+ \pi^-, \Sigma \pi}+ \frac{1}{2}  t_{\Sigma^0 \pi^0, \Sigma \pi}|^2 |\vec{k}|^3 |t_T|^2,
\end{equation}
\end{widetext}
with 
\begin{subequations}
\begin{align}
& |\vec{p}_a| = \frac{\lambda^{\frac{1}{2}}(s, M_p^2, M_p^2)}{2 \sqrt{s}}, \\
& |\vec{p}_1| = \frac{\lambda^{\frac{1}{2}}(s, M_p^2, M_{\text{inv}}^2)}{2 \sqrt{s}}.
\end{align}
\end{subequations}
This last step is important to account for the phase space
of this process
that depends on $|\vec{p}_1|$, which is tied to $M_{\rm inv}$.

Finally, we should integrate out the $\cos \theta$ in
Eq.~\eqref{eq:finalmassint}
but $C'$ in $C''$ depend on it.
The resultant factor of the $\cos \theta$ integration
is
denoted by $C'''$
and since we do not know the expression for $C'$,
we take $C'''=1$.
This means that from now on we will use arbitrary units
(a.u.) for the cross section.

Thus, we end up with
\begin{widetext}
\begin{equation}
\frac{d^2 \sigma_{pK^+\pi\Sigma}}{dM_{\text{inv}} d m_{\text{inv}}}=\frac{C''' 2 |\vec{p}_a| |\vec{p}_1|}{|M_{\text{inv}}-M_{N^*} + i \frac{\Gamma_{N^*}}{2}|^2} |\vec{\tilde{p}}_{\pi}| |t_{\Sigma^+ \pi^-, \Sigma \pi}+ \frac{1}{2}  t_{\Sigma^0 \pi^0, \Sigma \pi}|^2 |\vec{k}|^3 |t_T|^2.
\end{equation}
\end{widetext}

\section{Results}
\begin{figure}[t]
 \includegraphics[width=9cm]{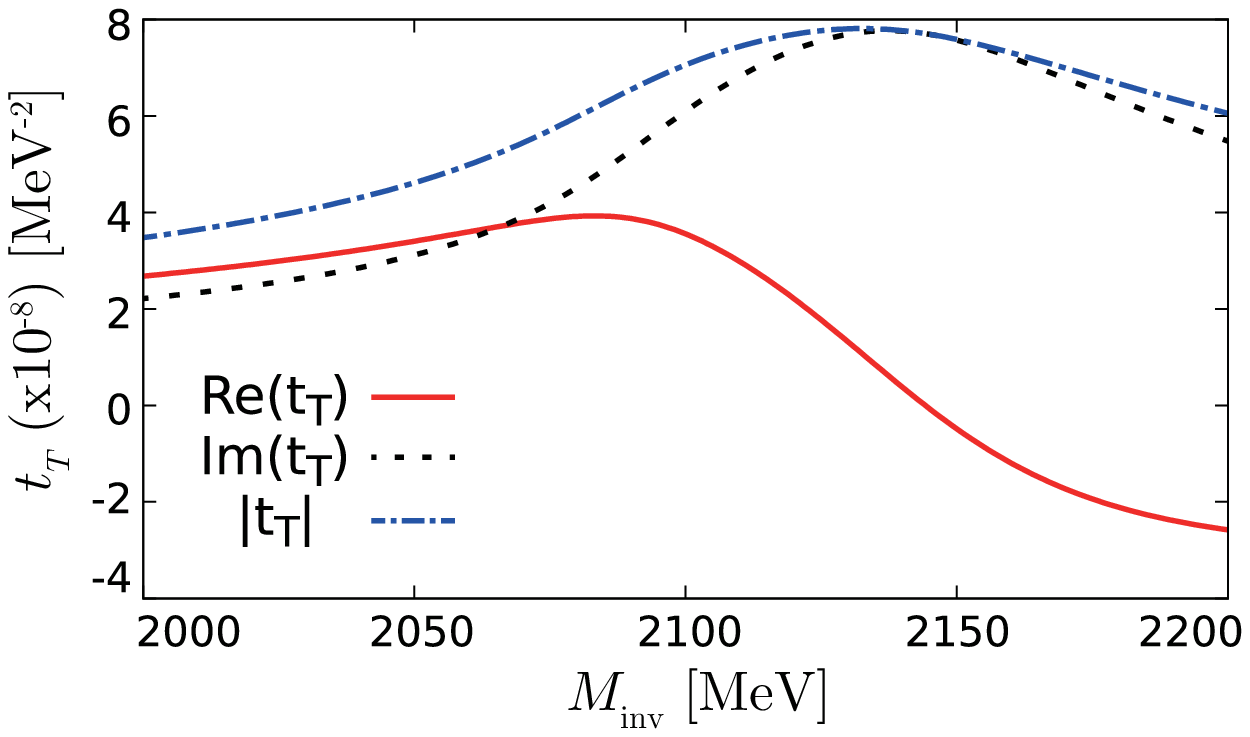}
 \caption{Re($ t_T $), Im($ t_T $) and $ \vert t_T \vert $ of Eq. (\ref{eq:tsing_int}).}
 \label{figtT}
\end{figure}

In Fig.~\ref{figtT}, we show the real, imaginary part and absolute
value of the amplitude $ t_T $ of Eq. (\ref{eq:tsing_int}) as a function
of
{the invariant mass of the $K\Lambda(1405)$, $M_{\rm inv}$,}
by fixing the invariant mass of $\pi\Sigma$, $m_{\rm inv}$, at
$1400$~MeV.
The absolute value of $t_T$ has a peak around $2140$~MeV as expected
from the condition for a triangular singularity by Eq.~(18) of
Ref.~\cite{Bayar:2016ftu},
{and the peak is dominated by the
imaginary part of the amplitude.
As mentioned in Ref.~\cite{Sakai:2017hpg}, the peak of the imaginary
part is responsible for the triangle singularity.}

\begin{figure}[t]
 \includegraphics[width=9cm]{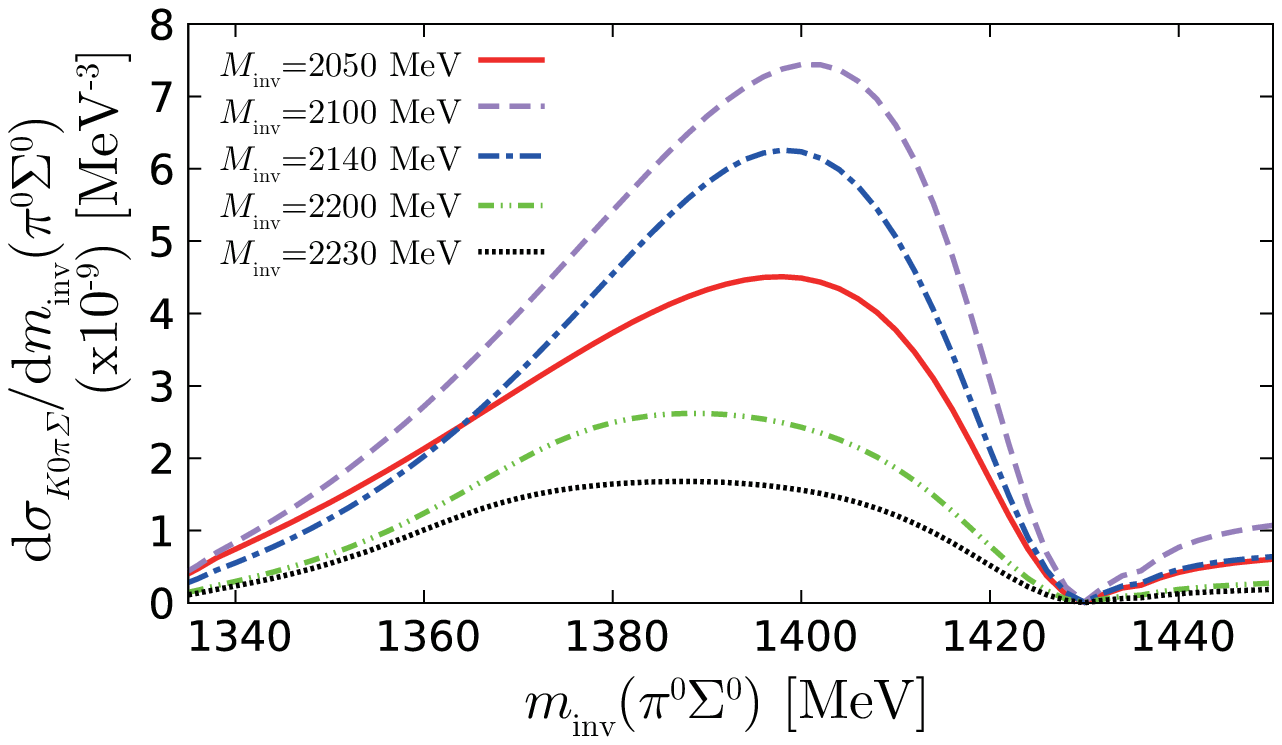}
 \includegraphics[width=9cm]{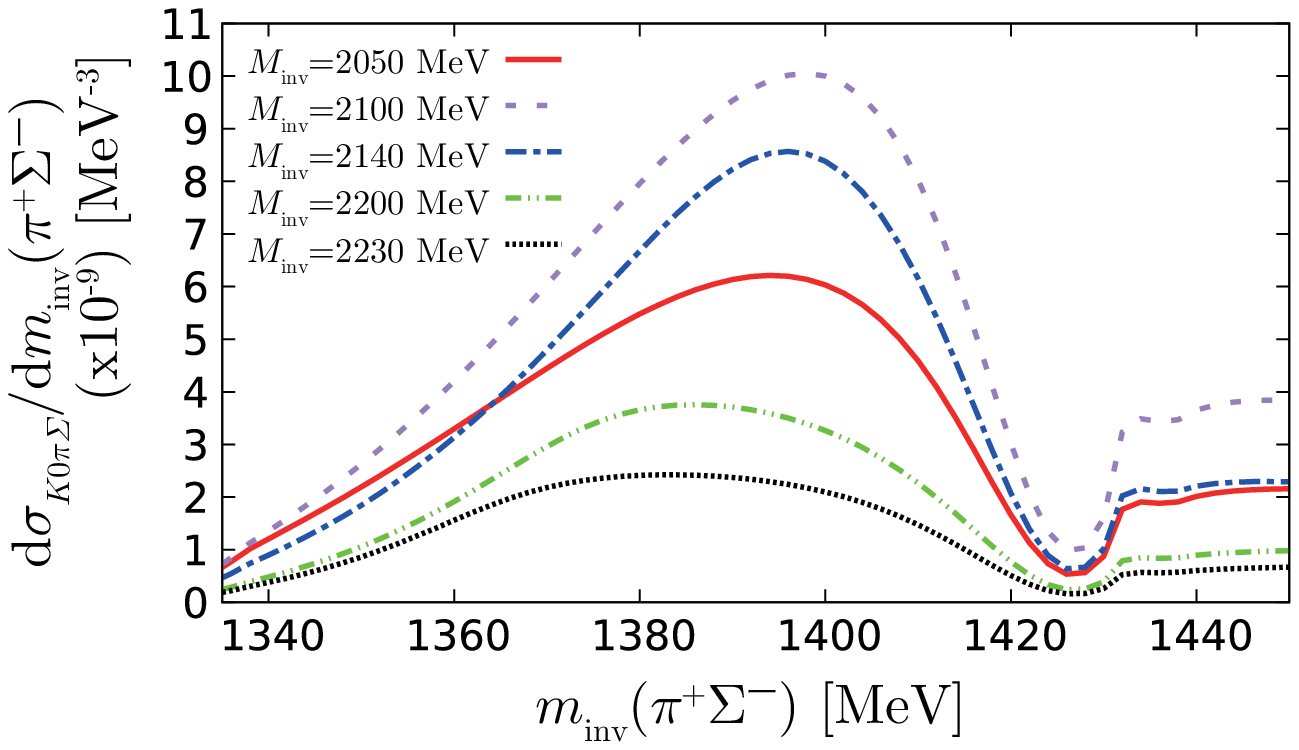}
 \includegraphics[width=9cm]{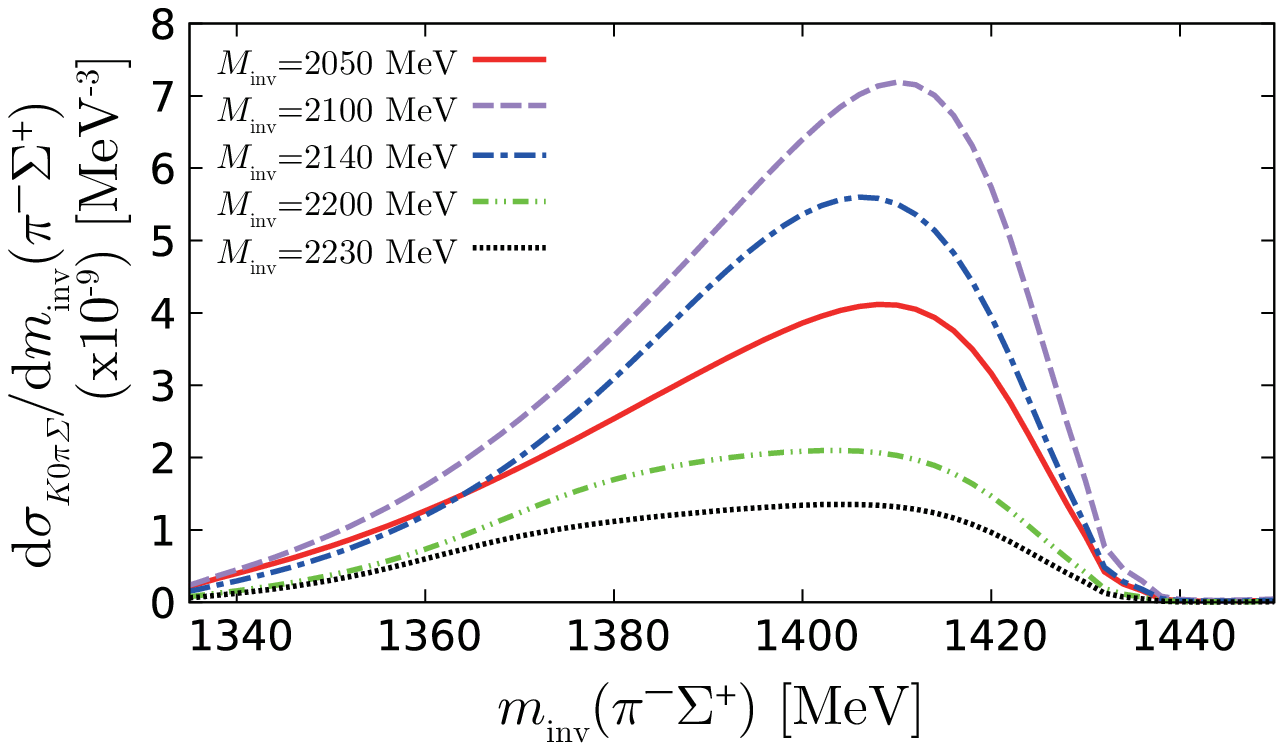}
 \caption{The $ \dfrac{d \sigma}{dm_{\rm inv}} $ mass distribution as a
 function of $ m_{\rm inv}(\pi^0 \Sigma^0)  $,  $ m_{\rm inv}(\pi^+
 \Sigma^-) $ and $ m_{\rm inv}(\pi^- \Sigma^+) $ for the $\pi^- p
 \rightarrow K^{0} \pi \Sigma$ scattering with several fixed values of
 $M_{\rm inv}$.}
 \label{figpi0S0}
\end{figure}

In Fig.~\ref{figpi0S0}, we plot the mass distribution of the  $\pi^-p\rightarrow K^0\pi\Sigma$
scattering process as a function of $ m_{\rm inv}(\pi^0 \Sigma^0) $, $
m_{\rm inv}(\pi^+ \Sigma^-) $ and $ m_{\rm inv}(\pi^- \Sigma^+) $
with a fixed {of}
{$\sqrt{s}=M_{\rm inv}$}
$=2050, ~2100,~ 2140, ~ 2200,~2230 $~MeV.
Let us first look at the $ \pi^0 \Sigma^0 $ mass distribution in
Fig.~\ref{figpi0S0}.
{At $M_{\rm inv}=2140$~MeV, where a peak associated with the
triangle singularity is expected from the formula in
Ref.~\cite{Bayar:2016ftu},}
we can see a clear peak at $1400$~MeV
{associated with
$\Lambda(1405)$ in the $\pi\Sigma$ invariant mass.}
As we see in the figure
{the largest strength is obtained with $M_{\rm inv}=2100$~MeV.}
A peak is found around $1385$ MeV for  $ M_{\rm inv} =2200, ~2230$ MeV
with a smaller strength, and the peak position moves towards
{higher energy a little}
for $ M_{\rm inv}=2050$~MeV.
{In the case of the $\pi^+\Sigma^-$ and $\pi^-\Sigma^+$ final state, while the
basic features are shared with the $\pi^0\Sigma^0$,
the
peak positions of the the $\pi^+\Sigma^-$ mass distribution}
are about $5$~MeV
less than that
of the $\pi^0\Sigma^0$ mass distribution, and
the peak positions in the $ \pi^- \Sigma^+ $ mass distribution are about
$5$~MeV bigger than the values of the $\pi^0 \Sigma^0 $ mass
distribution with a similar width and strength.
{Among these processes, the $\pi^+\Sigma^-$ gives the largest
strength.
This is roughly because the $t_{\Sigma^-\pi^+,\Sigma\pi}$ term is twice
larger than $t_{\Sigma^0\pi^0,\Sigma\pi}$ in Eq.~(\ref{eq:t_finals1}).}

\begin{figure}[t]
 \includegraphics[width=9cm]{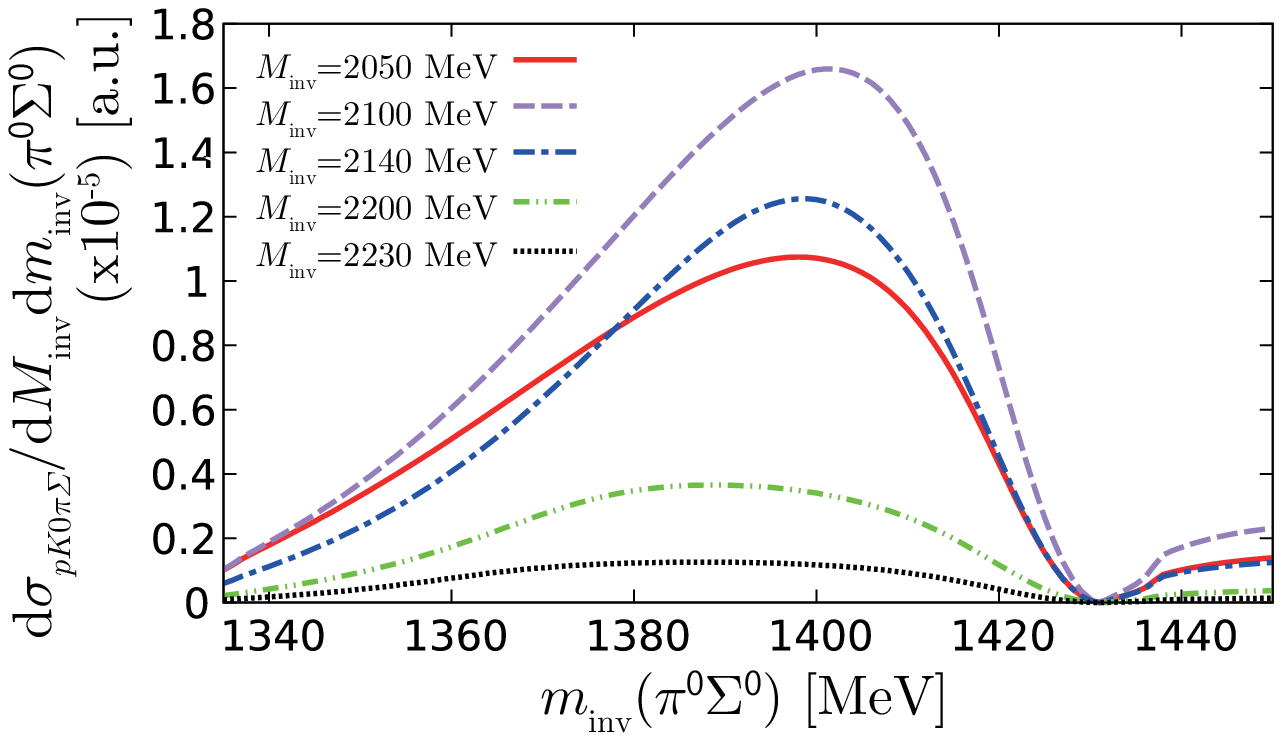}
 \includegraphics[width=9cm]{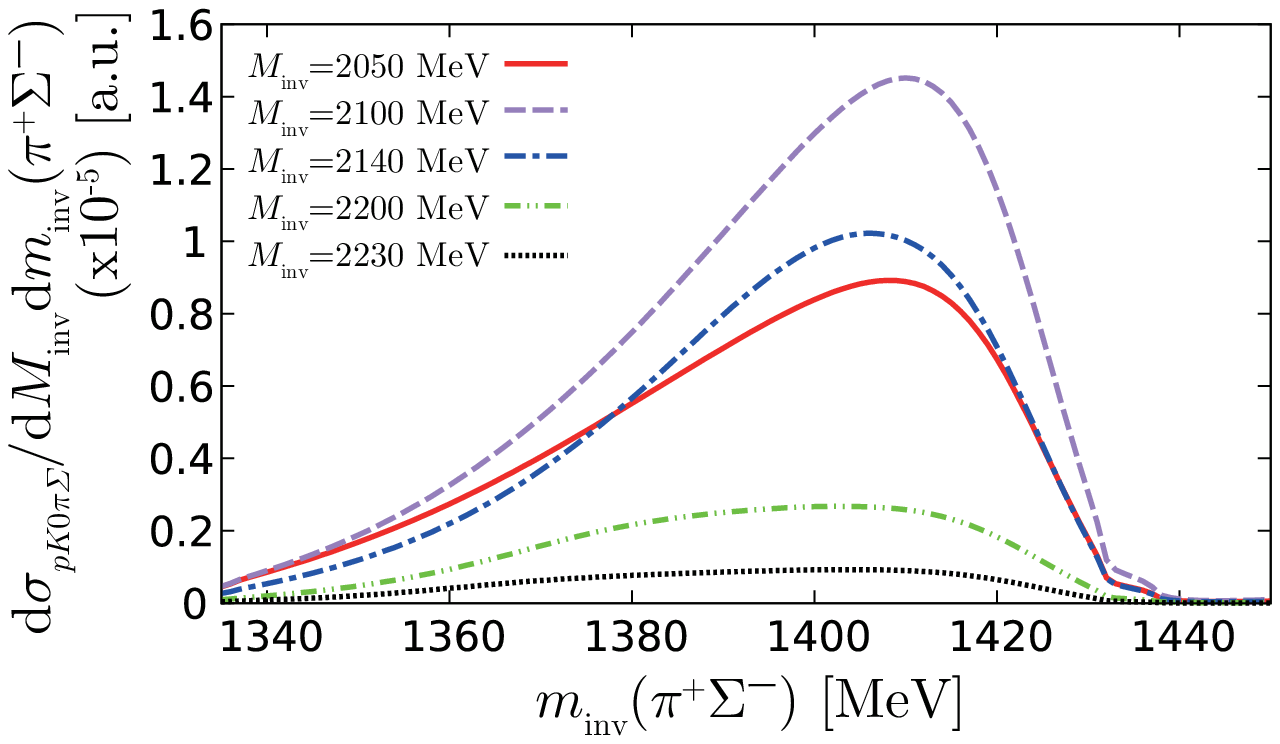}
 \includegraphics[width=9cm]{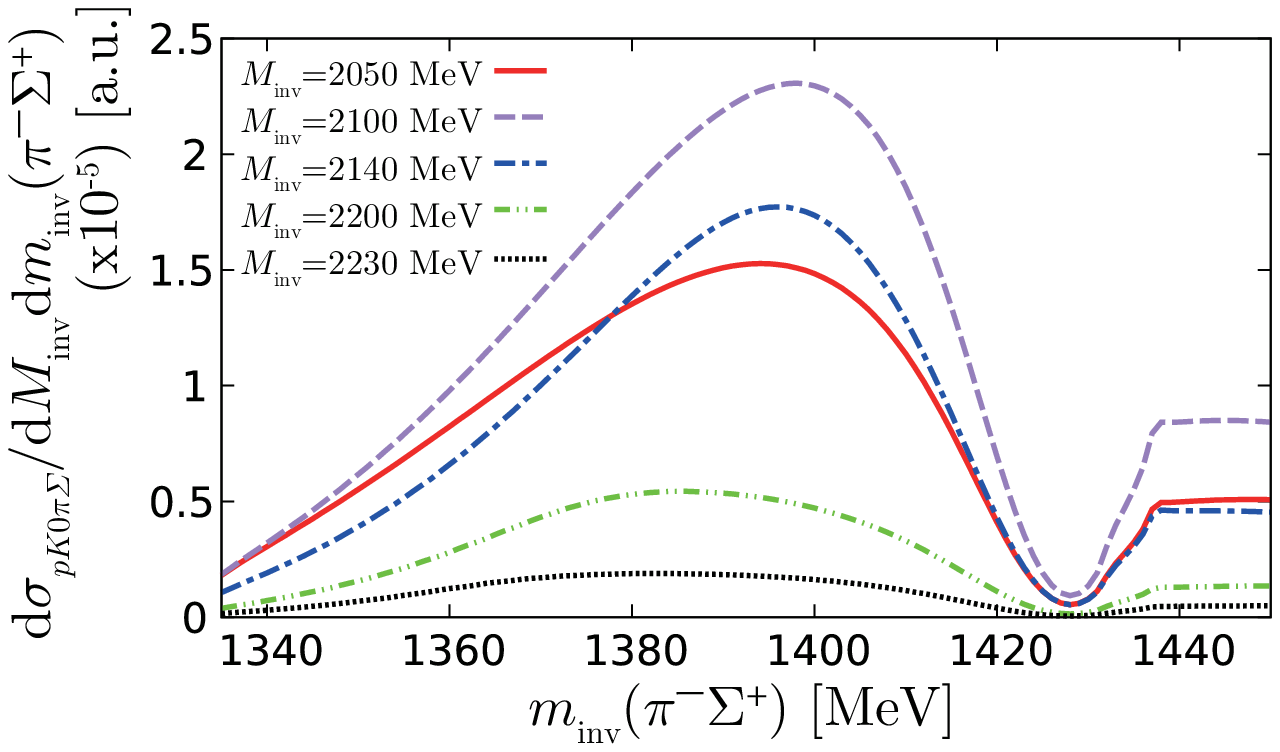}
 \caption{The $ \dfrac{d^2 \sigma_{pK^+\pi\Sigma}}{dM_{\rm inv} dm_{\rm inv}} $ as a
 function of $ m_{\rm inv}(\pi^0 \Sigma^0)  $,  $ m_{\rm inv}(\pi^+
 \Sigma^-) $ and $ m_{\rm inv}(\pi^- \Sigma^+) $  for the $p p
 \rightarrow p K^{+} \pi \Sigma$ scattering with several fixed values of
 $M_{\rm inv}$
 and $\sqrt{s}=3179$~MeV.}
 \label{figpi0S0PP}
\end{figure}

In Fig. \ref{figpi0S0PP}, we show the results of $ \dfrac{d^2
\sigma_{pK^+\pi\Sigma}}{dM_{\rm inv} dm_{\rm inv}} $ for the $p p
\rightarrow p K^+\pi \Sigma$ scattering as a functions of $ m_{\rm
inv}(\pi^0 \Sigma^0)$, $m_{\rm inv}(\pi^+ \Sigma^-)$ and $m_{\rm
inv}(\pi^-\Sigma^+)$, respectively.
{The total energy of the system $\sqrt{s}$ is fixed at 3179~MeV
which can be accessed
experimentally
\cite{Zychor:2007gf,Agakishiev:2012xk,Adamczewski-Musch:2016vrc}.}
{The dependence on $m_{\rm inv}$ is similar to that in
$d\sigma_{K^0\pi\Sigma}/dm_{\rm inv}$.}
In the case of the $ \pi^0 \Sigma^0 $ the peak is located at
$1400$~MeV  by fixing $ M_{\rm inv}=2140$ MeV. For $ M_{\rm inv}=2200$ and
$2230$~MeV, the peak positions move towards
$1380$ MeV and also the
widths are broader than that of the  $1400$ MeV case. Decreasing the
value of $ M_{\rm inv}$ to $2100$ MeV, we obtain the peak position
around $1405$ MeV. The shape of results are similar for the $ \pi^+
\Sigma^- $ and $ \pi^- \Sigma^+ $ mass distributions, but the peak
positions are $10$ MeV bigger for the case of $ \pi^+ \Sigma^- $  and
$5$ MeV smaller for $ \pi^- \Sigma^+ $ mass distribution.
In these processes, the $\pi^-\Sigma^+$ gives the largest
strength because of the additional factor two for the
$t_{\Sigma^+\pi^-,\Sigma\pi}$ term in Eq.~(\ref{eq:finalmassint})
compared with $t_{\Sigma^0\pi^0,\Sigma\pi}$. We should note that the peak with 
this mechanism appears at lower $\pi \Sigma$ invariant mass than with the model of \cite{genglam}, where 
the peak showed at $1420$ MeV. This is due to the fact that with the TS 
the $\Lambda(1405)$ is formed by $\pi \Sigma$, rather than $\bar{K} N$, and this channel couples
mostly to the lower mass state of the two $\Lambda(1405)$ states \cite{Jido:2003cb}.

\begin{figure}[h]
 \includegraphics[width=9cm]{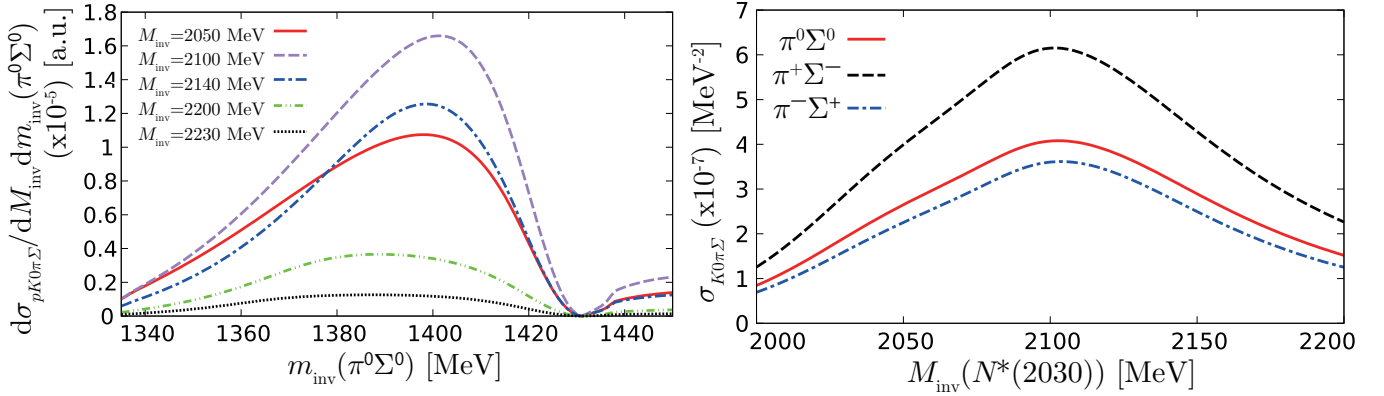}
 \caption{The cross section of the $\pi^-p\rightarrow K^0\pi\Sigma$
 process $\sigma_{K^0\pi\Sigma}$ as a function of $ M_{\rm inv} $  for
 the $\pi^- p \rightarrow K^{0} \pi \Sigma$ scattering.
 The red solid line corresponds to the
 $ \pi^0 \Sigma^0 $, the black dash line  the $ \pi^+ \Sigma^- $ and the
 blue dash-dot line $ \pi^- \Sigma^+ $.}
 \label{figpiPsigma}
\end{figure}

For the case of the $\pi^-p\rightarrow K^0\pi\Sigma$ reaction,
we integrate $ \dfrac{d \sigma_{K^0\pi\Sigma}}{dm_{\rm inv}} $ over
$ m_{\rm inv}$ in the range of the $ \Lambda(1405) $ peak,
{$m_{\rm inv}\in(m_{\pi}+m_{\Sigma},
1450~{\rm MeV}),$ with $m_\pi$ and $m_\Sigma$ the isospin-averaged mass of
$\pi$ and $\Sigma$,} and we obtain the cross section of $\pi^- p \rightarrow
K^{0} \pi \Sigma$, $\sigma_{K^0\pi\Sigma}$, as a function of  $ M_{\rm inv} $.
The results are represented in Fig.~\ref{figpiPsigma}. There
are peaks around $2100$ MeV for all cases.
though the expected value of triangular singularity is $2140$ MeV.
This is because the $N^*$ resonance in the $K^* \Sigma$
production has a peak around $2030$~MeV  (the term
$1/|M_{\text{\rm inv}}-M_{N^*} + i \frac{\Gamma_{N^*}}{2}|^2$ in
Eq.~(\ref{eq:finalmassint})).

\begin{figure}[t]
 \includegraphics[width=9cm]{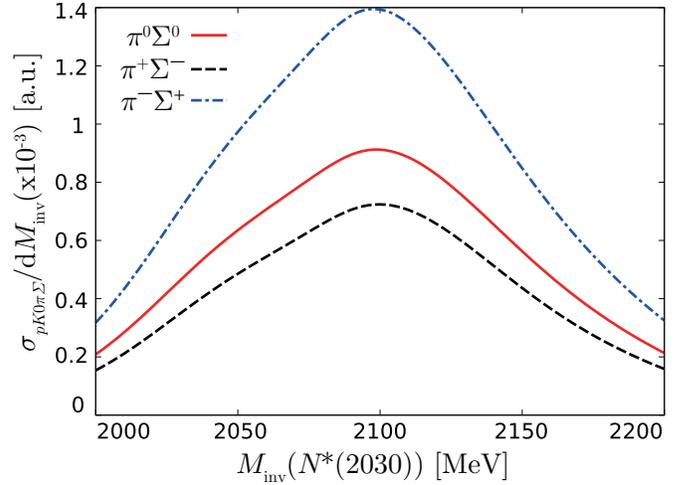}
 \caption{The $ \dfrac{d \sigma_{pK^+\pi\Sigma}}{d M_{\rm inv}} $  as a function of $
 M_{\rm inv} $ for the $p p \rightarrow p K^{+} \pi \Sigma$ scattering
 with fixed value of $ \sqrt{s} =3179$~MeV. The red solid line
 corresponds to the $ \pi^0 \Sigma^0 $, the black dash line  the $ \pi^+
 \Sigma^- $ and the blue dash-dot line $ \pi^- \Sigma^+ $.}
 \label{figPPsigma}
\end{figure}
For the case of the $pp\rightarrow pK^+\pi\Sigma$ reaction,
integrating now the $ \dfrac{d^2 \sigma_{pK^+\pi\Sigma}}{dM_{\rm inv} dm_{\rm inv}} $
over the $ m_{\rm inv} $  we obtain $ \dfrac{d \sigma_{pK^+\pi\Sigma}}{d M_{\rm inv}}$
which are shown in Fig. \ref{figPPsigma} as a function of  $ M_{\rm inv}
$ for $ \pi^0 \Sigma^0 $, $ \pi^+ \Sigma^- $ and $ \pi^- \Sigma^+
$. Similarly we get peaks around $2100$ MeV for the three cases. 

{In the $\pi^-p$ and $pp$ reactions, the strength is largest for the
$\pi^+\Sigma^-$ and $\pi^-\Sigma^+$ final state, respectively,
reflecting the strength before the integration shown in
Figs.~\ref{figpi0S0} and \ref{figpi0S0PP}.}

\section{Summary}
We have carried out a study of contributions of a triangle diagram to
the the $\pi^-p\rightarrow K^0\pi\Sigma$ and $pp\rightarrow
pK^+\pi\Sigma$ processes.
In both reactions, the triangle diagram is formed by the $N^*$ decaying
first to $K^*$ and $\Sigma$, the $K^*$ decays into  $\pi K$, and then the
$\Sigma$ and the $\pi$  merge to give  $\Lambda(1405)$, which finally
decays into $\pi\Sigma$.
In this process, the $K^*\pi\Sigma$ loop generates a triangle singularity
around $2140$~MeV in the invariant mass of $K\Lambda(1405)$ from the
formula of Eq.~(18) of Ref.~\cite{Bayar:2016ftu}. 
We evaluate the real, imaginary part and absolute value of the amplitude
$ t_T $ and find a peak around $ 2140 $ MeV.
We calculate  the $ \dfrac{d \sigma_{K^0\pi\Sigma}}{dm_{\rm inv}} $
with some values of $M_{\rm inv}$ in the $\pi^-p\rightarrow K^0\pi\Sigma$ reaction and $ \dfrac{d^2
\sigma_{pK^+\pi\Sigma}}{dM_{\rm inv} dm_{\rm inv}}$ with some
values of $M_{\rm inv}$ and fixed $\sqrt{s}=3179$~MeV
as a function of $ m_{\rm inv}(\pi^0 \Sigma^0) $, $ m_{\rm inv}(\pi^+
\Sigma^-) $ and $ m_{\rm inv}(\pi^- \Sigma^+) $.
In these distributions, we see  peaks
around $1400$ MeV, representing clearly  the $\Lambda(1405)$.
Integrating over the $ m_{\rm inv} $ we obtain $\sigma_{K^0\pi\Sigma}$
and $ \dfrac{d \sigma_{pK^+\pi\Sigma}}{d M_{\rm inv}} $ and these
distributions show a clear peak for $M_{\rm inv}(N^{*}(2030))$
around $ 2100 $ MeV.
The peak of the singularity shows up around $2140$~MeV. This peak position of the
triangular singularity is lowered by the initial $N^*$ resonance peak
around $2030$~MeV in the $K^*\Sigma$ production.

Thus, our results constitute an interesting prediction of the triangle singularity effect
in the cross sections of these decays.
The work done here could explain why in the experiments of
Refs.~\cite{Agakishiev:2012xk,Adamczewski-Musch:2016vrc} the invariant
mass distribution of $\pi\Sigma$ for the $\Lambda(1405)$ are found at
lower invariant masses than in other reactions.
It would also be interesting to see if the predictions done here
concerning the triangle singularity are fulfilled by the experimental
data, an issue that has not been investigated so far.
This work also can serve as a warning to future experiments that
measure these interactions, that they should be careful when associating peaks in this energy region to resonances.

\section*{Acknowledgements}
R.P.~Pavao wishes to thank the Generalitat Valenciana in the program
Santiago Grisolia.
This work is partly supported by the Spanish Ministerio de Economia y
Competitividad and European FEDER funds under the contract number FIS2014-57026-REDT,
FIS2014-51948-C2-1-P, and FIS2014-51948-C2-2-P, and the Generalitat
Valenciana in the program Prometeo II-2014/068.

\bibliographystyle{plain}

\end{document}